\documentclass[aps,pra,groupedaddress,reprint,floatfix]{revtex4-1}
\usepackage{graphicx}
\usepackage{amsmath}

\begin{document}
\title{Collisional trap losses of cold, magnetically-trapped Br atoms}
\author{J. Lam}
\affiliation{Department of Chemistry, University of Oxford, Chemistry Research Laboratory, Mansfield Rd, Oxford, OX1 3TA, United Kingdom.}
\author{C. J. Rennick}
\affiliation{Department of Chemistry, University of Oxford, Chemistry Research Laboratory, Mansfield Rd, Oxford, OX1 3TA, United Kingdom.}
\author{T. P. Softley}
\email{tim.softley@chem.ox.ac.uk}
\affiliation{Department of Chemistry, University of Oxford, Chemistry Research Laboratory, Mansfield Rd, Oxford, OX1 3TA, United Kingdom.}
\date{\today}

\begin{abstract}

Near-threshold photodissociation of Br$_2$ from a supersonic beam produces slow bromine atoms that are trapped in the magnetic field minimum formed between two opposing permanent magnets.
Here, we quantify the dominant trap loss rate due to collisions with two sources of residual gas: the background limited by the vacuum chamber base pressure, and the carrier gas during the supersonic gas pulse.
The loss rate due to collisions with residual Ar 
in the background
follows pseudo first-order kinetics, and the bimolecular rate coefficient for collisional loss from the trap is determined by measurement of this rate as a function of the background Ar pressure.
 This rate coefficient is smaller than the total elastic collision rate coefficient, as it only samples those collisions that lead to trap loss, and is determined to be $\langle\nu\sigma\rangle = (1.12\pm0.09)\times10^{-9}\,\text{cm}^3\, \text{s}^{-1}$.
The calculated differential cross section can be used with this value to estimate a trap depth of $293\pm24\,\text{mK}$.
Carrier gas collisions occur only during the tail of the supersonic beam pulse.
Using the differential cross section verified by the background-gas collision measurements provides an estimate of the peak molecular beam density of $(3.0\pm0.3)\times10^{13}\,\text{cm}^{-3}$ in good agreement with the prediction of a simple supersonic expansion model.
Finally, we estimate the trap loss rate due to Majorana transitions to be negligible, owing to the relatively large trapped-atom phase-space volume.

\end{abstract}


\maketitle

\section{Introduction}

The availability of a source of cold halogen atoms offers the opportunity to broaden the study of low-temperature chemical dynamics.
At very low temperatures the de Broglie wavelength 
of the collision partners 
becomes larger than the length scale of reaction energy barriers, enabling the ability to observe resonances associated with tunneling in chemical reactions at cold and ultracold temperatures\cite{Balakrishnan1,Balakrishnan2}.
For example, it is due to this effect below \mbox{10 K} that the cross sections for the reaction of F with H$_2$($\nu=0$ $j=0$) to form HF ($\nu' = 0,\, 1,\, 2$) are calculated to increase with decreasing collisional energy\cite{FH2Weck}.
A cold source of halogens would also be of interest for application  in the study of low-temperature ion-molecule collisions; for example cold  Br atoms
 could be combined with a cold source of BrCl$^+$, prepared in a single quantum state and sympathetically cooled into a Coulomb crystal of Ca$^+$ cations in a linear Paul trap.
This would be a good system to study ion-molecule chemistry because the very long lifetimes of both the BrCl$^+$ ($1.7\times 10^7$~s and $6.9\times 10^6$~s in its rotational ground and first excited state at 10~K, respectively\cite{willitschBrCl}) and the non-polar Br$_2^+$ would facilitate state-to-state kinetics measurements.

The generation of a cold source of halogen atoms by Doppler cooling is not currently possible however:  
the lowest single-photon transition of the halogens lies in the vacuum ultraviolet, a spectral region for which sufficiently powerful continuous laser systems for Doppler cooling are not widely available\cite{LaserVUVWalz}.
We have previously demonstrated  the production of cold Br atoms in the photodissociation of Br$_2$\cite{WillPCCP}, and confinement of the ground-state atoms in a permanent-magnet trap for durations up to 99~ms\cite{Rennick2014}. Permanent magnetic traps can provide larger trap depths for confinement of cold atoms than those produced by non-superconducting electromagnetic coils \cite{PermanentMagTollett}, which 
are limited to a trap depth of few thousand gauss.
The larger depth allows for the trapping of atoms and molecules with small magnetic moments, or higher kinetic energy, and hence obviates the need for additional Doppler cooling of the atoms.
While the photodissociation  method would, in principle, allow reloading and accumulation of trapped atom density at the 10~Hz experimental repetition rate,   various losses limit the accumulation and the ultimate density, and hence limit the applicability of this source to collisional studies.

In this paper, we quantify three mechanisms for loss of trapped atoms: elastic collisions with atoms in the tail of the supersonic beam pulse, elastic collisions with residual gas in the trap chamber, and Majorana transitions to anti-trapped states at the magnetic-field minimum. The characterization reported here facilitates the design of second-generation setups with the appropriate properties for accumulation of density of the trapped atoms.





\section{Experimental}\label{sec:experiment}
\subsection{Photodissociation}
\begin{figure}[htb]
    \centering
    \includegraphics[width=0.5\textwidth]{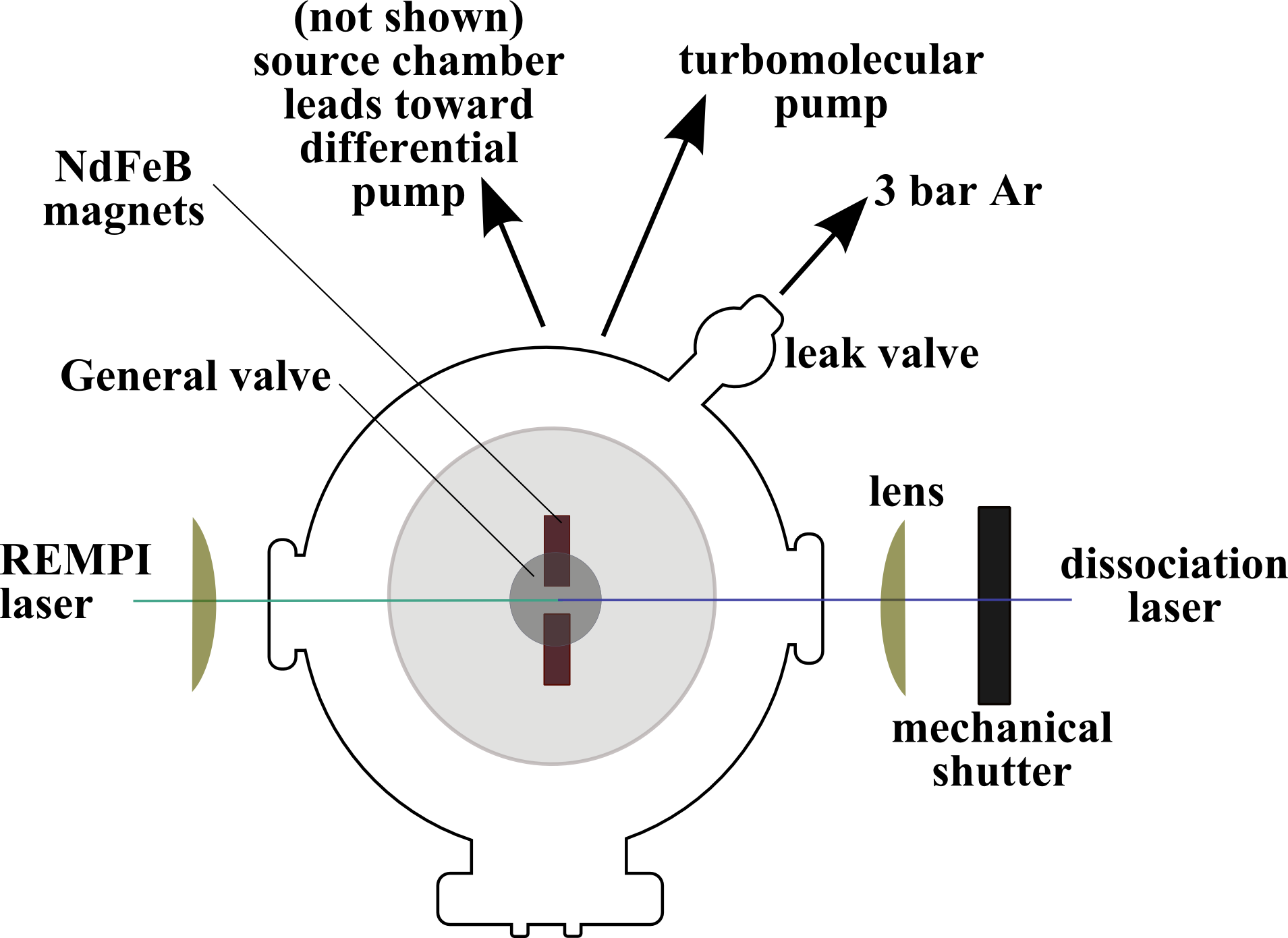}
    \caption{(Color online) Diagram of the experimental setup looking along the molecular beam axis. The differentially pumped source chamber of the skimmed supersonic beam is not shown, but lies below the trapping chamber.}
    \label{fig:experimentalsetupzoomout}
\end{figure}

A diagram of our experimental setup is shown in Fig. \ref{fig:experimentalsetupzoomout}. 
A pulsed, skimmed, supersonic molecular beam of 12\% Br$_2$ seeded in Ar is formed by a pulsed General Valve nozzle at a backing pressure of 2~bar.
A \mbox{460 nm} Nd:YAG pumped pulsed dye-laser saturates the $\text{B}^{3}\Pi^{+}_u \leftarrow X^{1}\Sigma^{+}_g$ transition of Br$_2$, leading to prompt dissociation. The angular distribution of the fragment recoil velocity peaks in a direction parallel to the laser polarization.
In the predominant dissociation channel,
each parent Br$_2$ molecule produces one ground state Br$({}^{2}\text{P}_{3/2})$ and one spin-orbit excited Br$({}^{2}\text{P}_{1/2})$ atom.
As shown previously, the center-of-mass frame recoil velocity of the Br fragments $u_{\text{Br}}$ is\cite{WillPCCP}
\begin{equation}
    u_{\text{Br}} = \sqrt{2 E^{\text{avg}}_{\text{kin}} \frac{m_{\text{Br}^*}}{m_{\text{Br}} m_{\text{Br}_2}}},
\label{equ:UVMI}
\end{equation}
\noindent
where $E^{\text{avg}}_{\text{kin}}$ is the average kinetic energy of the Br fragments, $m_{\text{Br}^*}$ is the mass of the excited state Br fragment, $m_{\text{Br}}$ is the mass of the ground state Br fragment (which may be different from the excited state fragment due to isotope effects), and $m_{\text{Br}_2}$ is the mass of the Br$_2$ parent molecule.
The recoil kinetic energy $E^{\text{avg}}_{\text{kin}}$  arises from the excess photon energy above the dissociation limit and is given by
\begin{equation}
    E^{\text{avg}}_{\text{kin}} = \mathit{h}\nu + E^{\text{Br}_2}_{\text{int}} - ( D^{\text{Br}_2}_0 + E^{\text{Br}}_{\text{int}} + E^{\text{Br}^*}_{\text{int}} ),
    \label{equ:KEVMI}
\end{equation}
where $\nu$ is the frequency of the laser, $E^{\text{Br}_2}_{int}$ and $D^{\text{Br}_2}_0$ are the internal and dissociation energies of Br$_2$, respectively, and $E^{\text{Br}}_{\text{int}}$ and $E^{\text{Br}^*}_{\text{int}}$ are the internal energies of ground and excited Br atoms, respectively. 

The center-of-mass velocity vector of the Br atoms traveling backwards along the molecular beam direction is canceled by that of the incoming molecular beam, so the atoms are stationary in the laboratory-frame.
The ground state Br$(^{2}\text{P}_{3/2})$ atoms are detected via $2+1$ resonance-enhanced multi-photon ionization (REMPI) at  226~nm\cite{BrREMPICooper}. The Br$^+$ cations produced by REMPI are extracted into a 50~cm long time-of-flight tube and detected by a multichannel plate (MCP). The current from the MCP is amplified and recorded on an oscilloscope.
At this wavelength, the probe laser has sufficient energy to dissociate a Br$_2$ molecule and ionize the resulting Br fragment.
The recorded signal therefore contains contributions from this one-color background, in addition to the two-color signal from the stopped Br atoms.
In order to distinguish the signal of interest, a mechanical shutter  is used to  block the photodissociation laser, and the one-color probe laser signal is recorded separately in alternation with the two-color signal.
The decay of the trapped Br atoms is mapped by delaying the probe laser with respect to the gas pulse \cite{WillPCCP}.
Ionization of the Br$_2$ parent molecules does not contribute to the recorded signal, as the wavelength used in the REMPI detection was far from resonance with the $2+1$ REMPI transition of Br$_2$ at 263~nm\cite{AshfoldBr2REMPI}.
The photodissociation process produces ground state Br atoms with a velocity distribution centered on zero.
These Br atoms are confined in all three axes by the magnetic minimum formed from two opposing NdFeB permanent magnets.

\subsection{Permanent magnet trap}

\begin{figure}[htb]
    \centering
    \includegraphics{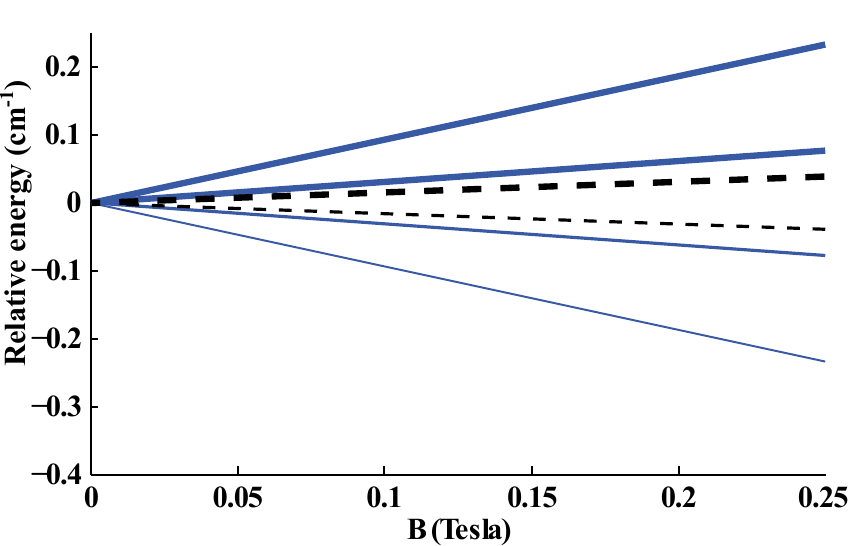}
    \caption{(Color online) Zeeman energies for the spin-orbit states of the excited $(j = 1/2)$ (black, dashed) and ground $(j = 3/2)$ (blue) fragments in a magnetic field, relative to field-free conditions. The bold lines with a positive gradient are the low-field seeking ($m_j = +3/2, +1/2$) states that can be trapped.}
    \label{fig:zeeman}
\end{figure}

In the presence of a magnetic field, the ground and spin-orbit excited states of the Br atom split into six Zeeman energy levels, labeled by the angular momentum $j$ and its projection on the axis defined by the field, $m_j$.
The states with energy that increase with increasing magnetic field ($m_j = +3/2,\,+1/2$) are termed low-field seeking, shown as bold lines in Fig.\ \ref{fig:zeeman}.
The Br atoms in these low-field seeking states are trapped.
As the $m_j$ sub-levels of the ground state show a larger increase in energy for a given field, a greater number of Br atoms in the ground state than excited state will be trapped and thus it is only these ground state Br atoms that we will examine in this paper.

\begin{figure}[htb]
    \includegraphics{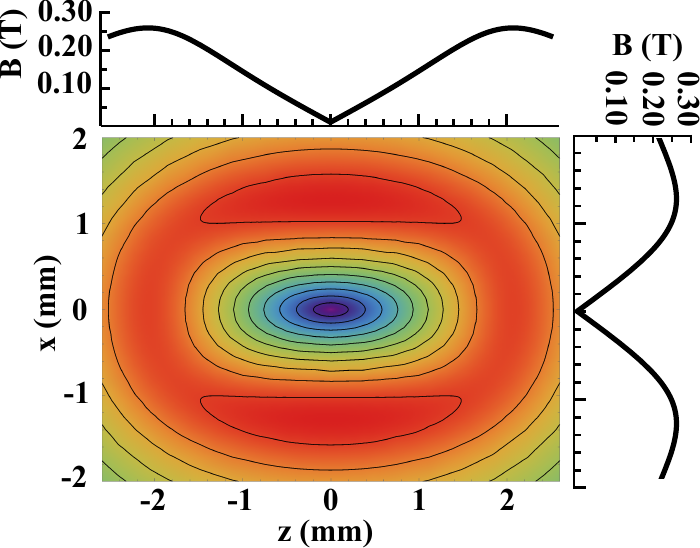}
    \caption{(Color online) Contour plot of the magnetic field strength in the plane perpendicular to two opposing bar-shaped magnets. The field minimum is shaded blue at $(z,x)=(0,0)$~mm, and rises to a maximum in all directions; contour lines are plotted at 25~mT intervals.
The trapping volume is constrained by the saddle points lying along the $x$ and $z$ axes -- illustrated by the line profiles along $x=0$~mm and $z=0$~mm -- and the field maximum at the magnet poles along the $y$ axis.}
\label{fig:trapField}
\end{figure}

The field surrounding a bar-shaped magnet with a magnetization close to saturation is given in an analytical form \cite{analyticalFields}. Given that the field due to one magnet at the position of the second magnet is 0.1~T, and ND48 neodymium-iron-boron magnets are highly resistant to demagnetization, we take the field in the gap 
between the magnets 
as the sum of the contributions from each magnet. 
The trapping field in the $(x,\,z)$ plane (perpendicular to the axis of the two magnets) is shown in Fig.\ \ref{fig:trapField}. The field magnitude increases linearly to a limit of 1.4~T in the direction towards each magnet face, but the trap depth is limited by a pair of saddle points along the $x$ and $z$ axes. 
The $z$ saddle point is lowest at approximately 0.24~T, calculated from the manufacturer's stated remanence, and we take this value as the predicted trap depth.
The maximum kinetic energy of a trapped bromine atom corresponding to this magnetic field corresponds to a temperature of 240~mK.

\subsection{Extraction of the trapped atom signal from the background}

\begin{figure}[htb]
    \centering
    \includegraphics[width=0.5\textwidth]{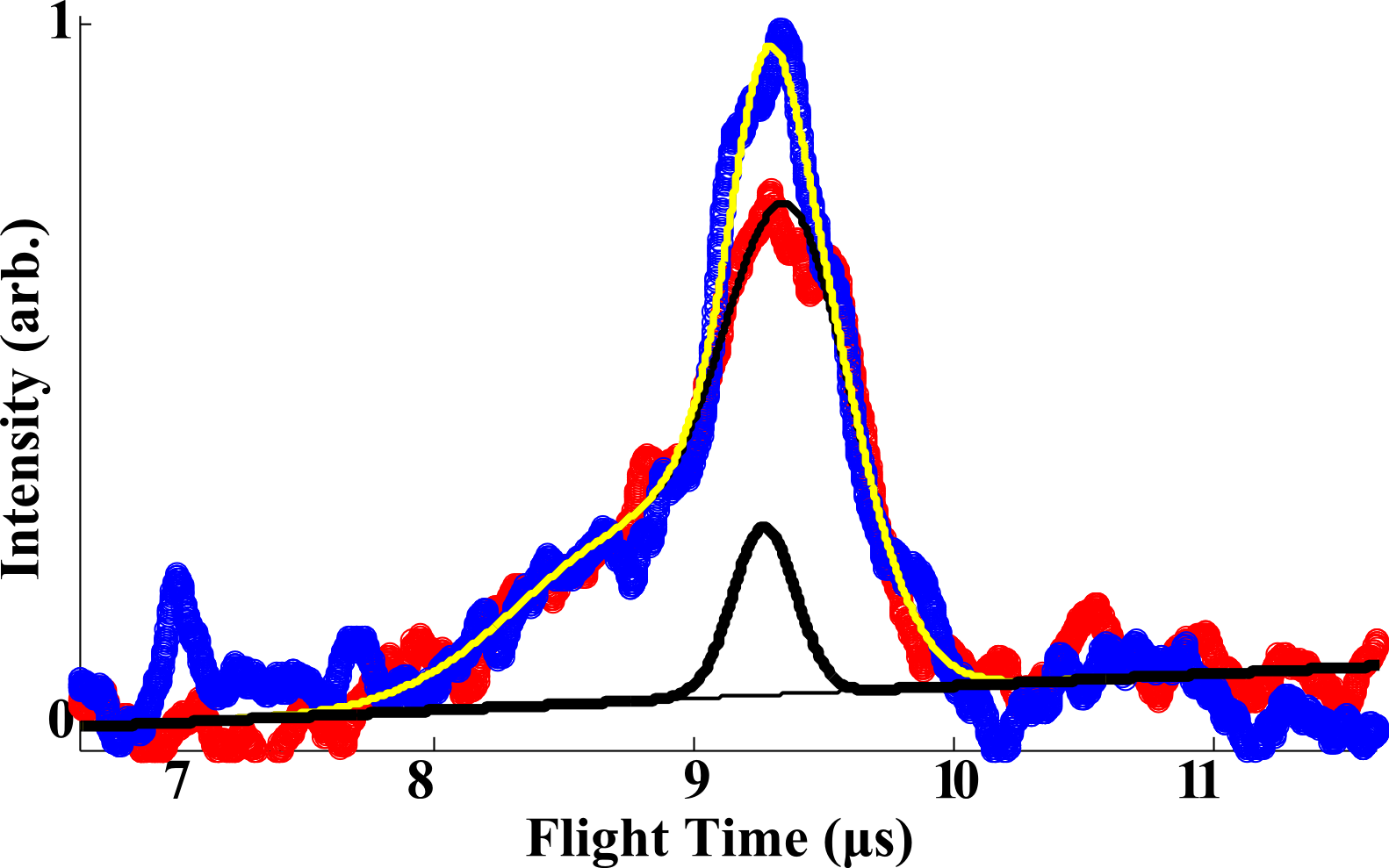}
    \caption{(Color online) Recorded oscilloscope traces from one- and two-color experiments fit by an ion-flight-time model. The one-color background is shown in red, and the two-color signal trace is shown in blue. The thick black curve illustrates the component resulting from trapped Br atoms.}
    \label{fig:mcmcsignal}
\end{figure}

Figure \ref{fig:mcmcsignal} shows the averaged one-color signal recorded with the dissociation laser blocked, and the two-color signal that includes a contribution from trapped Br. 
The recorded ion signal is directly correlated to the absolute number of Br$^+$ atoms formed by REMPI. 

The low count number means that the statistics of signal intensity follow a Poisson distribution, so regular $\chi^2$-distribution fitting is not statistically valid and fails to extract small signals from the large one-color background. 
Subtracting one Poisson distribution from another does not result in another Poisson distribution but rather a Skellam distribution, producing an apparently negative signal.
For sufficiently large signals, the measurement uncertainty is closely approximated by a normal distribution.
As the difference between two normal distributions is also a normal distribution, subtracting the background from the signal is sufficient.
Here, an \emph{additive} approach is taken, as the sum of two Poisson distributions is a Poisson distribution.

The integrated trapped atom density is extracted from the measured signal by fitting the one-color background and two-color signals to a time-of-flight model that represents the velocity-dependent arrival time distribution.
The model parameters are optimized by a Markov-chain Monte Carlo (MCMC) method using a maximum likelihood estimator described in Appendix \ref{sec:mcmc}.

The one-color background time-of-flight signal $\mu_{\text{bg}}(t)$ is represented by the sum of two Gaussian functions, representing the flight time of the two velocity components (backwards and forwards along the beam path) formed on dissociation and ionization:
\begin{equation}
	\mu_{\text{bg}}(t)= \omega + \sum\limits_{b=1}^2 \alpha_{b} \exp\left\{\frac{-(t - \beta_\text{b})^2}{2\gamma_\text{b}^2}\right\},
\label{equ:MLE1}
\end{equation} 
\noindent
where $\omega$ is a background offset that accounts, e.g., the electrical pick-up from the laser flash lamps, and $t$ is the flight time. 
The parameters $\alpha_\text{b}$, $\beta_\text{b}$, and $\gamma_\text{b}$ are the maximum intensity, flight time, and standard deviation, for each of the two Gaussian functions.
These six parameters are optimized to fit the data.

The two-color signal is represented by the background fit function in equation \eqref{equ:MLE1}, plus a single additional Gaussian function that represents the flight time for ions produced for trapped atoms:
\begin{equation}
\mu_\text{sig}(t) = \mu_\text{bg}(t) + \alpha_\text{s} \exp \left\{\frac{-(t - \beta_\text{s})^2}{2\gamma_\text{s}^2}\right\},
\label{equ:MLE2}
\end{equation}
where three additional parameters, $\alpha_\text{s}$, $\beta_\text{s}$, and $\gamma_\text{s}$, represent the maximum intensity, time-of-flight location, and standard deviation of the trapped Br signal respectively.
The six background parameters are kept fixed, and only the set of three additional parameters are adjusted in the fit.
This additive approach prevents the noise enhancement that occurs on simply subtracting the background from the signal traces and fitting a single Gaussian function to the resulting difference, and accounts correctly for the underlying statistics.
The average one- and two-color oscilloscope traces, overlaid with the MCMC-optimized flight time model are shown in
Fig.\ \ref{fig:mcmcsignal}. 

\section{Background gas collisions}\label{sec:bgcol}

In a  fraction of the elastic collisions between the trapped atoms and background gas,
 sufficient energy is transferred to overcome the trapping potential and to eject the atom.
The collision rate, and hence loss rate, is proportional to the background pressure.
The trap depth is determined by measuring this loss rate as a function of chamber pressure, and modeling the loss by pseudo first-order kinetics. 

The time-dependent trapped-atom density is probed by delaying the REMPI laser pulse with respect to the dissociation laser pulse.
A variable leak valve admits argon to the trap vacuum chamber setting the density of Ar collision partners.

\begin{figure}[htb]
    \centering
    \includegraphics{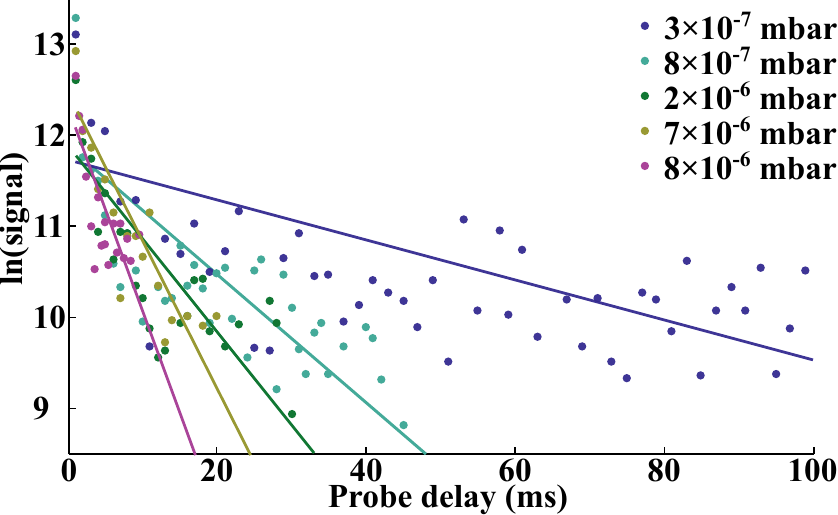}
    \caption{(Color online) Trapped Br atom signal as a  function of time probe delay for a range of chamber background pressures. The solid lines are fits to a pseudo-first order kinetic model of trapped atom density.}
    \label{fig:pressureDecays}
\end{figure}

\begin{figure}[htb]
    \centering
    \includegraphics{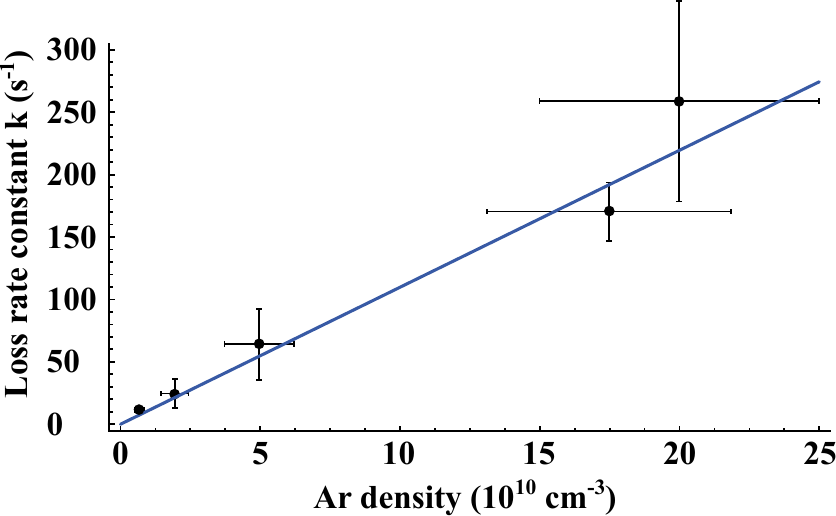}
    \caption{(Color online) Pseudo-first order trap loss rate coefficient as a function of background Ar density.}
    \label{fig:lossRate}
\end{figure}

For a given chamber pressure, the background Ar density is constant, and the trapped Br atom loss rate is well-represented by a pseudo-first order process expressed as
\begin{equation}
    \frac{d[\text{Br}]}{dt} = - k[\text{Br}],\label{equ:firstorder}
\end{equation}
\noindent
where $k = \langle v\sigma\rangle [\text{Ar}]$ is a pseudo first-order rate coefficient that depends on argon density and the velocity-weighted cross section for trap-loss collisions.
The integrated form of equation \eqref{equ:firstorder} fits the experimental data,  as shown by the  solid lines in Fig.\ \ref{fig:pressureDecays}, producing an estimate of $k$ for each argon background pressure.
Figure \ref{fig:lossRate} plots this rate constant as a function of Ar density, estimated from the measured chamber pressure assuming ideal gas behavior. The lowest pressure shown corresponds to the base pressure in the trapping chamber with the supersonic beam operating, i.e. with no extra gas admitted via the leak valve.

The gradient of the plot in Fig.\ \ref{fig:lossRate} gives an estimate of the velocity-weighted cross section $\langle v \sigma\rangle=(1.12\pm0.09)\times10^{-9}\,\text{cm}^{3}\,\text{s}^{-1}$, which is equivalent to the bimolecular rate coefficient.
This is the rate coefficient for collisional trap loss, and depends on the Br-Ar elastic collision cross section and the depth of the trap.
We now make use of this measurement together  with the differential cross section, calculated from the Br-Ar interaction potential, and establish the magnetic trap depth.

The Br-Ar interaction potential has been determined from crossed molecular beam collision experiments in an energy range comparable to collisions of thermal argon with trapped bromine\cite{Casavecchia1981-Interaction-potentials-for-Br2PAr}.
The strong spin-orbit coupling means that the electron orbital and angular momenta remain coupled throughout the collision, and the coupling between the adiabatic states of the Br-Ar molecular system is weak.
The collision can therefore be modelled purely elastically on each of the adiabatic potential energy curves.
The centro-symmetric potential is weakly attractive, with a well-depth of about 100~K, and strong short-range repulsion. 
Consequently, the angular distribution of scattered atoms is strongly forward-peaked over a wide range of collision energies. [Note that any alignment of the Br electronic orbital angular momentum caused by the magnetic field is ignored in these calculations.]

The scattering wavefunction in a central potential is represented by a partial-wave expansion in a basis of Legendre polynomials $P_l(x)$; the differential cross section for a collision of energy $E$ is then given by
\begin{equation}
	\sigma(\theta; E) = \frac{1}{4\mu^2 E^2} \sum_l \left|(2l+1) S_{l}(E) P_l(\cos\theta)\right|^2,
\end{equation}
where $S_{l}(E)$ is the energy-dependent scattering matrix element, derived in Appendix \ref{sec:crossSection}, and $\mu$ is the reduced mass of the collision system.

\begin{figure}[htb]
    \centering
    \includegraphics{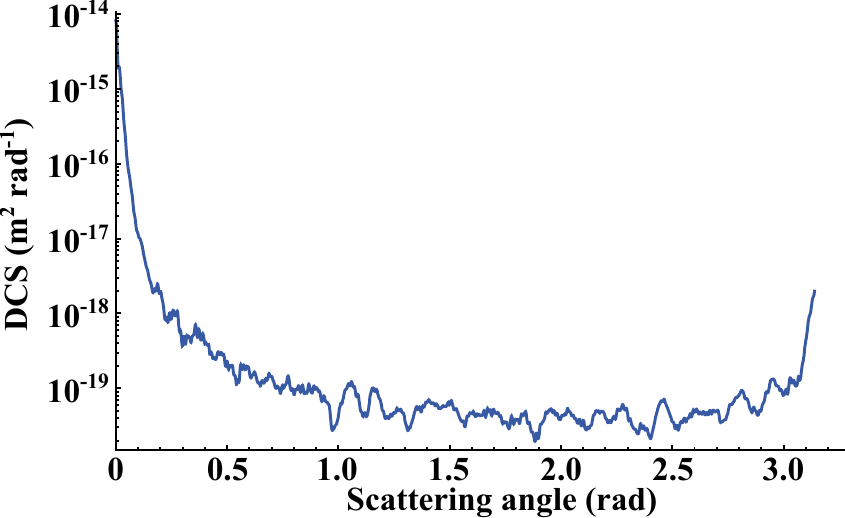}
    \caption{(Color online) Mean differential cross section calculated for collisions of room-temperature Ar atoms with trapped Br atoms, weighted over a 298~K Maxwell-Boltzmann velocity distribution.}
    \label{fig:dcs}
\end{figure}

The differential cross section $\sigma(\theta; E)$ is then weighted by the collision-energy distribution function $f(E)$, taken as a Maxwell-Boltzmann kinetic energy distribution
\begin{equation}
    \sigma(\theta) = \frac{2}{\sqrt{\pi}} \left( \frac{1}{k_{\text{B}}T} \right)^{3/2} \int_0^{\infty} \sigma(\theta; E) \sqrt{E} \exp\left\{-\frac{E}{k_{\text{B}}T}\right\}\, dE.
    \label{equ:vsigma}
\end{equation}
The differential cross section (DCS) averaged over the 298~K Maxwell-Boltzmann energy distribution is shown in Fig.\ \ref{fig:dcs}.

\begin{figure}[htb]
    \centering
    \includegraphics{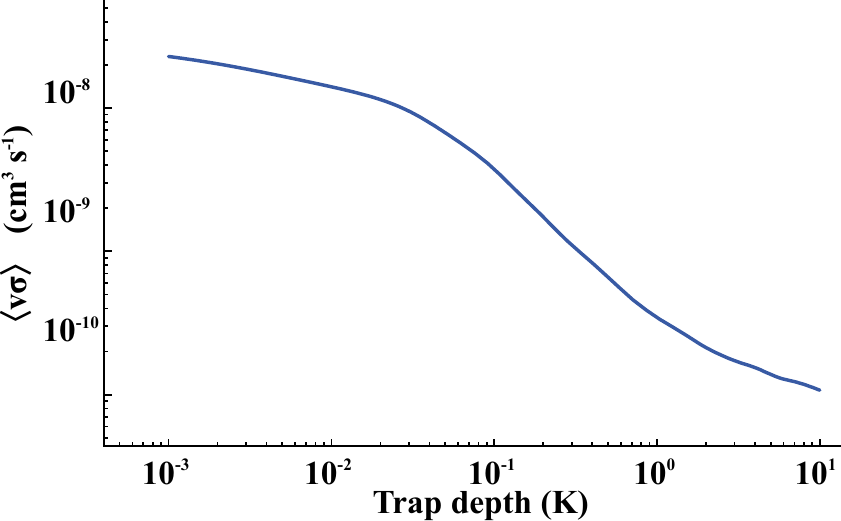}
    \caption{(Color online) Trap loss rate coefficient as a function of trap depth.}
    \label{fig:ratedepth}
\end{figure}

When the initial kinetic energy of the trapped atom is small enough to be neglected, the energy transferred in an elastic collision with relative velocity $v_r$ depends on the scattering angle $\theta$ as
\begin{equation}
	\Delta E = \frac{\mu^2}{m_2}\left|v_r\right|^2(1-\cos\theta),
\end{equation}
where $m_2$ is the mass of the trapped Br atom, and $\mu$ is the reduced mass of Br and Ar. 
For a single collision to eject an atom from the trap it must increase the kinetic energy of the trapped atom to overcome the trapping potential $U_{\text{trap}}$. 
This defines a minimum scattering angle $\theta_{\text{min}}$:
\begin{equation}
	\theta_{\text{min}} = \arccos\left\{ 1-\frac{m_2 U_{\text{trap}}}{\mu^2 \left| v_r \right|^2} \right\}.
\end{equation}
\noindent
The velocity-weighted cross section (or rate coefficient) obtained by integrating the DCS over this depth-dependent scattering angle, is shown in Fig.\ \ref{fig:ratedepth} as a function of the trap depth.
The experimentally-determined trap-loss rate coefficient  of $(1.12\pm0.09)\times10^{-9}\,\text{cm}^{3}\,\text{s}^{-1}$ therefore predicts a trap depth of $293\pm24\,\text{mK}$, which 
is in reasonable agreement with the estimated value of $240\pm24\,\text{mK}$
 calculated by magnetic field simulations using the manufacturers specified magnetization \cite{Rennick2014} (see Section IIA).

\section{Molecular beam collisions}\label{sec:mbcol}

During the earliest period of trap loading the dominant loss mechanism is collisions with the supersonic beam carrier gas.
The beam has a similar average speed compared to the room-temperature Maxwell-Boltzmann distribution, but with velocity directed predominantly along a single axis.

Collisions between the beam and trapped atoms lead to loss with a similar probability as collisions with the background gas, but at a higher rate owing to the increased on-axis density.
The amount of trapped bromine remaining at the end of the supersonic beam pulse depends on two competing factors that are both determined by the timing of the dissociation laser pulse. On the one hand, the initially trapped Br signal should increase with the instantaneous $\text{Br}_2$ density at the time the dissociation laser is fired, and therefore is maximized by photodissociating at the peak of the beam pulse. On the other hand, producing the atoms at the peak of  the pulse may result in greater losses from collisions with the cumulative argon density in the tail of the beam, compared to the losses when photodissociation occurs later in the pulse.

\begin{figure}[htb]
    \centering
    \includegraphics{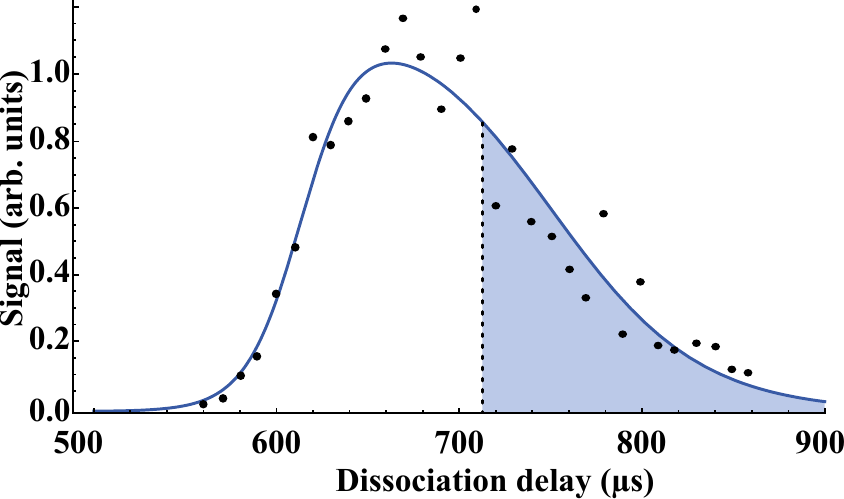}
    \caption{(Color online) Instantaneous molecular beam density determined from $\text{Br}_2$ photodissociation and ionization yield as a function of dissociation delay time relative to molecular beam valve trigger. The solid line is a fit to an exponential double-sigmoid function. The shaded region indicates integration range from the dissociation time to the end of the pulse, used in equation \eqref{equ:mbloss}.}
    \label{fig:beamprofile}
\end{figure}

Figure \ref{fig:beamprofile} shows a plot of the relative density of the  beam as a function of the delay between the valve trigger and the dissociation laser pulse.
Here, the ionization laser is fired coincidentally with the dissociation laser, and the two-color ion signal is proportional to the instantaneous $\text{Br}_2$ density.
We also take this measurement to be representative of the temporal profile of the relative carrier gas density, assuming minimal velocity slippage with respect to the seed $\text{Br}_2$.
The rising and falling edges are fitted with an exponential double-sigmoid function, with a decay rate approximately 2.5 times that of the rising edge, and a full-width at half-maximum intensity of $136\,\mu\text{s}$.

Direct measurement of the trapped-atom loss rate during the molecular beam pulse is not possible as there are two additional contributions to the observed signal intensity.
The high $\text{Br}_2$ density during the molecular beam pulse yields a significant one-color REMPI signal as the probe laser has sufficient energy to dissociate a molecule and ionize the fragments.
In addition, the fast atoms produced in the dissociation are still leaving the probed region during the molecular beam pulse. Hence immediately after parent molecule dissociation, the atom-signal time dependence is a convolution of both trapped atom loss and escape of high-velocity atoms from the probe volume.

\begin{figure}[htb]
	\includegraphics{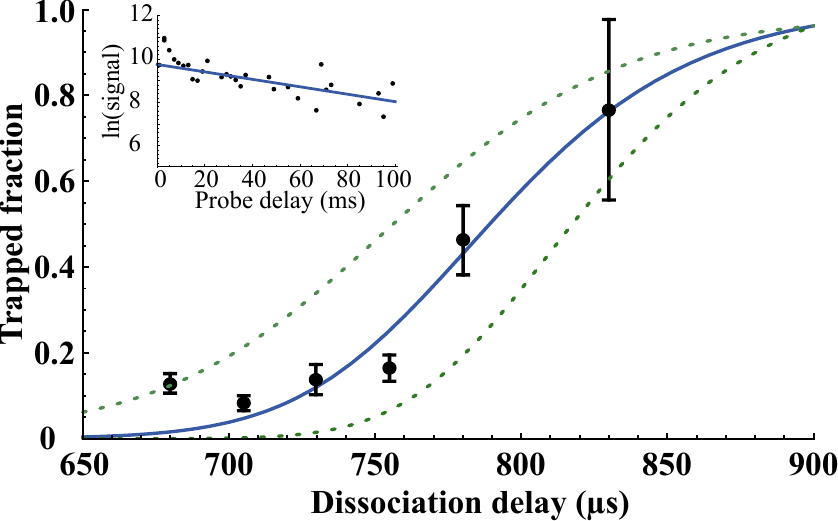}
    \caption{(Color online) Fraction of trapped atoms remaining after the molecular beam pulse, determined by extrapolation of trap loss curves. The inset shows an example of the extrapolation at a dissociation delay of $780\,\mu\text{s}$. The solid line is a fit to a kinetic model given by Eq.~\eqref{equ:mbloss}. The upper and lower dashed lines show the estimated fraction for half and twice the fitted Ar density respectively.}
	\label{fig:beamLossFit}
\end{figure}

We therefore use an indirect measure of initial Br atom density that only accounts for the trappable fraction.
The post-molecular beam signal decrease is fitted to an exponential decay, the rate of which is only dependent on the background pressure, as shown in section \ref{sec:bgcol}.
Extrapolating each of these fits back to \mbox{$t$ = 900 $\mu$s} gives an estimate of the trapped atom density at the end of the molecular beam gas pulse; this time is defined as the point at which the measured molecular beam intensity is indistinguishable from the background (Fig. \ref{fig:beamprofile}.)
This extrapolated value is then normalized by the molecular beam density at the instant of excitation to account for the $\text{Br}_2$ density during photodissociation (and hence the initial Br atom density).
Figure \ref{fig:beamLossFit} plots this relative trapped atom fraction as a function of dissociation time, and shows that exciting later in the molecular beam permits survival of a greater fraction of the original atom density.

The fractional trap loss during a molecular beam pulse is also modelled as a pseudo-first order process with a time-dependent rate coefficient arising from the time-dependent argon density, $k(t) = \langle v\sigma\rangle [\text{Ar}](t)$.
The rate coefficient $k=\langle v\sigma\rangle$ is then determined from the differential cross section for collisions at the $560\,\text{m s}^{-1}$ molecular beam velocity, integrated over the scattering angle limits determined from the trap depth in section \ref{sec:bgcol}.

Integrating the rate equation from the dissociation point $t_{\text{D}}$ to the end of the pulse at $t=900\,\mu\text{s}$ -- illustrated by the shaded region in Fig.\ \ref{fig:beamprofile} -- yields the fractional trap loss
\begin{equation}
    \frac{[\text{Br}](t=900\,\mu\text{s})}{[\text{Br}](t_\text{D})} = \exp \left\{ - \int_{t_{\text{D}}}^{t} [\text{Ar}](t') \langle v\sigma\rangle d t' \right\}.
	\label{equ:mbloss}
\end{equation}
The time-dependent Ar density is represented analytically by the amplitude-normalized double-sigmoid function, multiplied by the peak beam number density.
This equation is fitted to the data by a least-squares minimization of the peak argon beam density, and a scaling factor that accounts for the fraction of Br atoms produced with trappable velocities.
The solid line in Fig.\ \ref{fig:beamLossFit} demonstrates the fit for a peak argon density of $(3.0\pm0.3)\times10^{13}\,\text{cm}^{-3}$; the upper and lower dashed lines indicate the predicted dependence for a density a factor of two lower and higher respectively.

The intensity of a supersonic gas expansion (particles per steradian per second) is given by\cite{Beijerinck1981-Absolute-intensities,DePaul}
\begin{equation}
    I_0=1/8 f(\gamma) n_0 u_0 r^2,
\end{equation}
where $n_0$ is the source number density, $u_0$ is the source peak velocity, and $r$ is the radius of the nozzle. $f(\gamma)$ is a function of the heat capacity ratio:
\begin{equation}
    f(\gamma)=\sqrt{\frac{\gamma }{\gamma +1}} \left(\frac{2}{\gamma +1}\right)^{\frac{1}{\gamma -1}}.
\end{equation}
In the case of the mixture used here, gas properties are given by the mass-weighted mean of the components. The number density at the laser interaction region is then given by $n=\frac{I_0}{u_{\infty} d^2}$, where $u_{\infty}$ is the terminal velocity of the molecular beam, and $d$ is the distance from the nozzle to the interaction region.
This simple gas dynamic model predicts a peak argon beam density of $4.7\times10^{13}\,\text{cm}^{-3}$. This value is in good agreement with that determined experimentally in this work, especially taking into consideration that the orifice of the pulsed nozzle is not fully opened by the retraction of the sealing poppet, and therefore the effective nozzle diameter is slightly smaller than the actual one.
Evaluating the integral in equation \eqref{equ:mbloss} from the optimum photodissociation delay $t_{\text{D}}=775\,\mu\text{s}$ to the end of the pulse gives that 60\% of initially trapped atoms are removed by collision with the carrier gas under the experimental conditions used here.

\section{Trapped atom density}

Our trap depth estimated by background losses in Sec.~\ref{sec:bgcol} corresponds to a 7.8~ms$^{-1}$ maximum speed for trapped bromine atoms; atoms moving faster than this speed will eventually find the saddle point along the $z$ axis and leave the trap, regardless of their initial direction.
The trap  is conservative and thus provides no additional cooling; it only acts to filter the slowest velocities from the distribution produced on photodissociation.

\begin{figure}[htb]
    \includegraphics{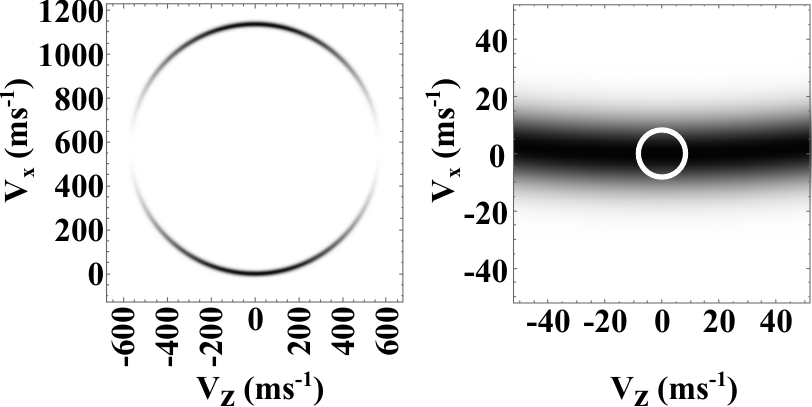}
    \caption{(Color online) Ground-state Br atom velocity distribution after photodissociation at $460$~nm, given by the equation in the text. The right-hand panel shows a zoom around $V_z=0$~ms$^{-1}$, with the maximum trappable velocity highlighted as a white circle.}
    \label{fig:vel}
\end{figure}

Photodissociation of Br$_2$ at 460~nm yields atoms in two product channels: a pair of ground-state Br atoms, and one ground-state and one spin-orbit excited atom.
The second channel produces trappable bromine atoms with a cylindrically-symmetric angular distribution relative to the laser polarization given in terms of the second Legendre polynomial $P_2$ as $I(\theta) = \sigma/(4\pi)(1+\beta P_2(\cos\theta)$ with a limiting value of $\beta=1.5$.
The velocity distribution of this product channel after photodissociation with the laser polarization parallel to the supersonic beam propagation is shown in Fig.\ \ref{fig:vel}.

Only a small portion of the atoms are formed with velocities that can be trapped, illustrated by the white circle in Fig.\ \ref{fig:vel}.
Integrating the velocity distribution within this bound reveals that $4.5\times10^{-3}$ of the yield in this product channel are trappable.
This channel represents 60\% of the total photodissociation yield, and the laser power is sufficient to excite 50\% of the molecules in the supersonic expansion. The argon-Br$_2$ mixing fraction is set by the vapor pressure of bromine to approximately 8\% at the backing pressure used here.

The focused photodissociation laser illuminates a 750~$\mu$m-diameter cylindrical column through the center of the trap.
When exciting at the optimum point of the gas pulse quantified in section \ref{sec:mbcol}, and accounting for the 60\% loss due to the tail of the gas pulse, a total of $5\times10^{5}$ Br atoms in the $m_j=3/2$ state are trapped initially.
The cloud of atoms expands to fill the trap, defined by the volume encompassed by the 0.24~T depth, to a final density of approximately $1\times10^{8}$~cm$^{-3}$.
This density is of the same order of magnitude as that attainable by many other trapping methods.

\section{Majorana losses}\label{sec:majorana}

For long trapping times, loss due to Majorana spin flips may need to be considered for magnetic traps. 
When the Br atom moves across a region with sufficiently small field, near the center of the magnetic trap, it can change from a low to a high-field seeking $m_j$ sub-level and is expelled from the trap. 
Majorana transitions may occur when the rate of change of the magnetic field experienced by the atom $\left|{d B}/{d t}\right|$ exceeds the Larmor frequency $\omega_L$\cite{GomerMajorana}.
Substituting the notional time dependence of the magnetic field for its spatial gradient and the velocity of a trapped atom yields the inequality that must be satisfied for a Majorana transition to occur:
\begin{equation}
    \omega_L = \frac{\mu_B B}{\hbar} \ll \left| \frac{d B}{d t}\right| \frac{1}{B} = \left| \frac{d B}{d r}\right| \frac{v_\text{Br}}{B},
    \label{equ:majorana1}
\end{equation}
where $B \equiv B(\textbf{r})$ is the magnetic field experienced by the trapped bromine atom, $\mu_\text{B}$ is the Bohr magneton, and $v_{\text{Br}}$ is the laboratory-frame velocity of the Br atoms.

The gradient of the quadrupole-like field in this trap is linear near the center, so the inequality in \eqref{equ:majorana1} depends most strongly on the local field magnitude $B(r)$, and trapped particle velocity $v_{\text{Br}}$.
The inequality is most likely to be satisfied, resulting in trap loss transitions, near the center, where $v_{\text{Br}}$ is maximum and $B$ is minimum.
Distinguishing such losses from those described previously is not experimentally possible here, so instead we estimate the contribution of Majorana losses from classical simulations of the atoms' motion in the trap.

\begin{figure}[htb]
	\centering
	\includegraphics{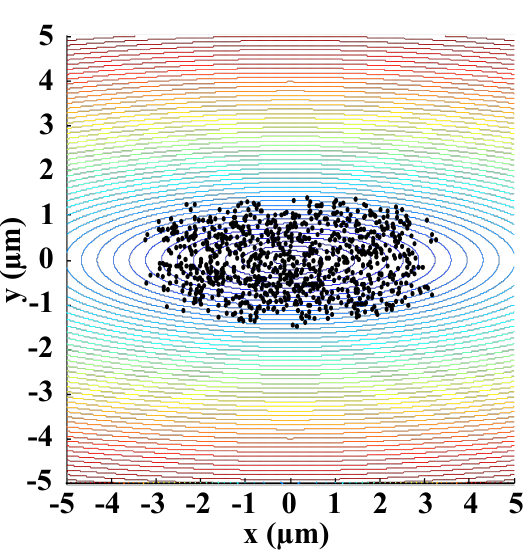}
    \caption{(Color online) Map of the locations near the center of the trap at which 200 out of 1350 simulated Br atoms will undergo a Majorana transition after 5 seconds. The contour lines qualitatively indicate the magnetic field magnitude near the center of the magnetic trap. The magnets are situated at $\pm1000\,\mu\text{m}$ in the $y$ axis and the lasers are parallel to the $x$ axis.}
	\label{fig:majTrap}
\end{figure}

The trapped atom trajectory is simulated by numerically integrating the equations of motion, including the effect of the trapping field.
The field around a single bar magnet is determined analytically\cite{analyticalFields}, and the field within the two-magnet trap is determined from the contribution from each magnet assuming minimal demagnetization.
At each half timestep, the position and velocity of each Br atom is calculated from a velocity-Verlet algorithm, using the acceleration due to the trapping field.
The Br atoms are then moved to their new positions, and each atom is then tested to see if its position and velocity satisfies the condition for Majorana transition to a high-field seeking state, resulting in trap loss.
This process is repeated for several thousand simulated particles with random initial positions and velocities, for a simulated time of 5~s.
Figure \ref{fig:majTrap} plots the location of each particle at the time it undergoes a transition and is lost from the trap.

This procedure provides a very conservative loss rate estimate of $0.3\,\text{s}^{-1}$, a rate much lower than that due to elastic collisions.
This is much slower than similar estimates of the Majorana loss rate from other magnetic traps owing to the much higher trapped atom phase space volume. For example the trapped atom velocity in this case has a maximum of $7.8~\text{m s}^{-1}$. 

\section{Conclusions}

In conclusion, we have examined the potential loss mechanisms of cold magnetically-trapped Br atoms.
The dominant loss is through elastic collisions with background gas exchanging sufficient kinetic energy to overcome the trapping potential.
The differential cross section for Ar-Br collisions is strongly forward-peaked, so the majority of collisions do not transfer sufficient energy and the trap loss rate is less than might be expected \emph{a priori}.
Measurement of the trap loss rate as a function of background gas density provides a direct measure of the rate coefficient  for these collisions.
Comparison with the calculated differential cross section allows an estimate of the trap depth.
Quantifying the equivalent loss rate for collisions with the argon carrier gas in the supersonic beam provides an estimate of the beam density.

The rate of loss due to collisions with the molecular beam can be fitted, using the verified differential cross section,  to a molecular beam density that is  in close agreement with that determined from gas dynamics.
Monte Carlo molecular dynamics simulations indicate that losses due to Majorana transitions are much slower and only significant at longer periods of confinement on the order of seconds.

Overall, under our experimental conditions, the fractional loss due to collisions with the molecular beam (60\%) in the first few hundred microseconds, is comparable to the losses due to collisions with the background gas over 100~ms at a pressure of $3 \times 10^{-7}$ mbar.
The conservative nature of the trapping potential means that it is not possible to remove the slow atoms from the path of the gas pulse.
Shortening the pulse duration of the supersonic beam could in principle reduce both collisions within the beam and also reduce the background pressure (and hence the collision rate with the background).
Fast-acting pulsed valves based on piezoelectric\cite{Irimia2009-A-short-pulse-7-mus-FWHM} or electromagnetic actuators are capable of producing gas pulses shorter than 10~$\mu$s, or mechanical shutters and rotating choppers could also be implemented to shorten the existing gas packet.
An evaluation of the available methods for implementation in this magnetic trap system is beyond the scope of the present paper.

Further work would be needed to examine other types of losses of Br atoms in our trap, which would become significant and observable over longer periods of time on the order of seconds.
One mechanism for further loss would be collisions with subsequent beam pulses (in our experiments these occur every 100 ms.) With a static magnetic trap it is not possible to move the trapped atoms away from the beam axis. 
At higher trap densities, inelastic collisions of trapped Br atoms with another trapped Br atom or with Ar background gas, might induce scattering loss. These processes are unlikely to be spin-orbit changing processes, given the high energy gap to the excited state (more than 10 times the thermal energy), but $m$-changing collisions would lead to trap loss.

The characterization of the Br atom loss mechanisms described in this paper will be very useful in future work toward realizing the accumulation of density of these Br atoms over timescales of the order of seconds. This work provides a step towards expanding the breadth of cold atomic species that can be trapped at high density to include the halogen atoms, which are of great interest for the study of cold chemical dynamics.

\section*{Acknowledgments}

T.P.S. acknowledges the financial support of the EPSRC under Grant EP/G00224X/1. J.L. is grateful for support from an AAUW American Fellowship, and C.J.R. from the Ramsay Memorial Trust.

\appendix
\section{Monte Carlo Markov chain using Maximum Likelihood Estimation (MLE)}\label{sec:mcmc}

Conventional $\chi^2$ fitting assumes that the error in a signal is normally distributed. 
This is not the case for small counting-based signals and background subtraction and fitting can fail when extracting small signals from large backgrounds.
The technique used here is derived from Bayesian inference, and seeks the most probable set of fit parameters given a set of observations\cite{MCMCGreen,MCMCHastings,MCMCVanDyke}.
The procedure is to define the distribution function for observation errors, derive a likelihood estimator function, and then maximize the likelihood value using a MCMC approach to sample the parameter space.

The recorded ion signal is proportional to the number of ions. The probability of observing $c$ counts, given the average expected intensity $\mu$, follows Poissonian statistics and can be expressed as
\begin{equation}
	p(c \mid \mu) = \frac{\mu^c}{c!}\exp\left\{{-\mu}\right\}.
	\label{equ:Poisson}
\end{equation} 

The likelihood estimator is defined in terms of the probability distribution function:
\begin{equation}
	\mathcal{L} = \prod_{i} p(c_\text{i} \mid \mu_\text{i}(\bar{a})),
	\label{equ:LE}
\end{equation} 
where the expected intensity in a given time bin $\mu_\text{i}(\bar{a})$ (i.e. the amplitude of a single point in the oscilloscope trace) now depends on a set of model parameters $\bar{a}$.
The product is over all observations, in this case  the entire oscilloscope trace.
It is these parameters we seek to optimise, and in the case described here are the set of Gaussian function parameters describing the amplitude, arrival time, and width of our signal; $\bar{a}_{\text{bg}} = [\alpha_{1},\, \beta_{1},\, \gamma_{1},\, \alpha_{2},\, \beta_{2},\, \gamma_{2}]$ for the 1-color background and $\bar{a}_{\text{sig}} = [\alpha_\text{s},\, \beta_\text{s},\, \gamma_\text{s}$] for the two-color signal.

The likelihood $\mathcal{L}$ is also known as the sampling distribution $p(c \mid \bar{a},\,\mathcal{I})$, which is the probability of observing $c$ counts, given parameters $\bar{a}$ and any prior information $\mathcal{I}$. Prior information includes things such as parameter constraints.
Bayes' theorem expresses this conditional probability in terms of the posterior probability we seek: $p(c \mid \bar{a},\, \mathcal{I})$ the probability of obtaining parameters $\bar{a}$, given a set of observations $c$ and prior knowledge $\mathcal{I}$\cite{BayesTheorem}
\begin{equation}
	p(\bar{a} \mid c,\, \mathcal{I}) = \frac{p(c \mid \bar{a},\, \mathcal{I})\,  p(\bar{a} \mid \mathcal{I})}{p(c \mid \mathcal{I})}.
	\label{equ:LE2}
\end{equation} 
The prior distribution $p(\bar{a} \mid \mathcal{I})$ represents known information about parameters $\bar{a}$ prior to observing $c$, and $p(c \mid \mathcal{I})$ is a normalizing constant that represents the unconditional distribution of $c$.

A Markov chain samples the posterior distribution $p(\bar{a} \mid c,\, \mathcal{I})$ over a series of iterations that converge toward a maximized value of the likelihood estimate $\mathcal{L}$ in equation \ref{equ:LE} \cite{MLEAtkinTutorial,MLEHannam}. 
At each iteration the value of $\mathcal{L}_{\text{current}}$ is calculated from the current set of parameters.
A new set of proposed parameters are selected at random from a probability distribution function centered around the value of the current set of parameters. 
The specific form of this selection probabilty distribution function does not affect the converged parameters, but is chosen to thoroughly sample parameter space.
Here a Gaussian probability distribution function was chosen to sample each parameter with a standard deviation of $10^{-3}$ of the parameter value as standard libraries exist for generating random numbers following this distribution.
The likelihood value for this new set of parameters is calculated, $\mathcal{L}_{\text{proposed}}$, and compared to the current likelihood:
\begin{equation}
    \alpha \leq \frac{\mathcal{L}_\text{proposed}}{\mathcal{L}_\text{current}},
	\label{equ:MLE4}
\end{equation} 
where $\alpha$ is a random number generated from a uniform distribution from 0 to 1 exclusive.
If this inequality is satisfied, then the current set of parameters is replaced with the new set, otherwise the current parameters are kept, and a new set of proposed parameters are generated.

In practice, $\log\mathcal{L}$ is typically used, rather than the direct value, as it aids numerical convergence:
\begin{align}
	\log\mathcal{L} &= \log\left(\prod_\text{i} p(c_\text{i};\mu_\text{i}(\bar{a}))\right)\\ 
		&= \sum\limits_{i}^n \left(-\mu_\text{i}(\bar{a})  + c_\text{i} \log\mu_\text{i}(\bar{a})\right).
	\label{equ:MLE3}
\end{align}
As $\mathcal{L}$ and $\log\mathcal{L}$ are monotonically related, maximizing this produces the same optimized parameters\cite{MLEMyungTutorial}.


Once the chain has converged, the accepted parameters $\bar{a}$ sample the posterior distribution, p($\bar{a} \mid c,\, \mathcal{I})$, and the mean and standard error of each parameter can be extracted from a sufficient number of iterations.

\section{Collision cross section for trapped atoms}\label{sec:crossSection}

The relative motion of the colliding atoms on each potential is described by the time-independent Schr\"odinger equation of the form
\begin{equation}
    \Psi''_l(r) = W_l(r)\Psi_l(r),\label{equ:schrodinger}
\end{equation}
\noindent
where $\Psi_l(r)$ is the radial wavefunction expanded in terms of partial waves. $W_l(R)$ is a coupling term that depends on the centrosymmetric interaction potential $V(r)$ including a centrifugal contribution as
\begin{equation}
    W_l(r)=\frac{2\mu}{\hbar^{2}}V(r) + \frac{l(l+1)}{r^{2}} - k^{2}.
\end{equation}
The collision wave vector $k$ depends on the collision energy $E$: $k^2=\frac{2\mu E}{\hbar^2}$.
A number of techniques have been developed to solve equations of the form of equation \eqref{equ:schrodinger}. In this simple, single-channel, case we use the log-derivative propagator using a constant reference potential\cite{Manolopoulos1986-log-derivative,Manolopoulos1987-reference-potential}.
The wavefunction is expanded as the sum of partial waves representing the incoming plane wave, and a spherical outgoing wave, in terms of the Legendre polynomials.

The log-derivative of the wavefunction is defined as $Y_l(r)=\Psi'_l(r)\Psi_l(r)^{-1}$. Differentiating and substituting the second derivative using \eqref{equ:schrodinger} yields the Ricatti equation $Y'_l(r) = W_l(r) - Y_l(r)^2$. This equation can not be integrated directly as the log-derivative is singular when the wavefunction vanishes, but can be solved across an interval ($a$, $b$) by the propagator
\begin{equation}
	Y_l(b) = \mathcal{Y}_4(a,\,b) - \mathcal{Y}_3(a,\,b)\left[Y(a) + \mathcal{Y}_1(a,\,b)\right]^{-1} \mathcal{Y}_2(a,\,b),
\end{equation}
where the propagator elements are
\begin{align}
    \mathcal{Y}_1(a,\,b)=\mathcal{Y}_4(a,\,b) = %
    &\left \{%
    \begin{aligned}
        |w|\text{coth}|w|h	&&	w^2\geq 0\\
        |w|\text{cot}|w|h	&&	w^2<0
    \end{aligned}%
    \right .\\ 
    \mathcal{Y}_2(a,\,b)=\mathcal{Y}_3(a,\,b) = %
    &\left \{%
    \begin{aligned}
        |w|\text{cosech}|w|h	&&	w^2\geq 0\\
        |w|\text{cosec}|w|h	&&	w^2<0
    \end{aligned}%
    \right .
\end{align}
In the case of this single-channel model, $w^2$ is equal to the coupling matrix $W_l(r)$ evaluated at $r=a$, and $h=b-a$. 
Starting from an initial value $Y_l(r_0)$, with $r_0$ well within the classical inner turning point, the integration proceeds by propagating out to $Y_l(b)$ in small steps using the mean interaction potential over the interval.

At the limit of large $r$ the usual asymptotic boundary condition for scattering problems is that the wavefunction can be written as the difference between incoming $I(r)$ and outgoing $O(r)$ waves with the collision energy dependent scattering matrix element for a particular partial wave $S_l(E)$
\begin{equation}
	\Psi_l(r\to\infty)\simeq I_l(r) - O_l(r) S_l(E).
    \label{equ:infWF}
\end{equation}
The values of $I_l(r)$ and $O_l(r)$ are determined by setting the potential to zero, such that
\begin{align}
	I_l(r) &= k^{-1/2}\hat{h}_l^-(kr),\\
		O_l(r) &= k^{-1/2}\hat{h}_l^+(kr),
\end{align}
\noindent
where $\hat{h}_l^-(kr)$ and $\hat{h}_l^+(kr)$ are the incoming and outgoing Riccati-Hankel functions respectively.
Substitution of the definition of the log-derivative wavefunction into \eqref{equ:infWF} yields a solution for the scattering matrix element:
\begin{equation}
	S_l(E) = \frac{Y_l(r_a)I_l(r_a) - I_l'(r_a)}{Y_l(r_a)O_l(r_a)-O_l'(r_a)},
\end{equation}
\noindent
where $r_a$ is some final value of the propagation radius in the asymptotic region of the potential.

\bibliography{ref}

\end{document}